\documentclass[sigconf]{acmart}
\AtBeginDocument{%
  }

\usepackage{amsfonts}
\usepackage{algorithmic}
\usepackage{textcomp}
\usepackage{xspace}
\usepackage{fontawesome}
\usepackage{multirow}
\usepackage{rotating}
\usepackage{soul}
\usepackage{makecell}
\usepackage{url}
\usepackage{enumitem}
\usepackage{pifont}
\usepackage{listings}

\newcommand{\lstbg}[3][0pt]{{\fboxsep#1\colorbox{#2}{\strut #3}}}
\lstdefinelanguage{diff}{
  basicstyle=\ttfamily\scriptsize,
  morecomment=[f][\lstbg{red!20}]-,
  morecomment=[f][\lstbg{green!20}]+,
  morecomment=[f][\textit]{@@},
}

\newcommand{\ie}{\emph{i.e.,}\xspace}
\newcommand{\eg}{\emph{e.g.,}\xspace}

\newcommand{\etal}{\emph{et~al.}\xspace}
\newcommand{\secref}[1]{Section~\ref{#1}\xspace}

\newcommand{\figref}[1]{Fig.~\ref{#1}\xspace}

\newcommand{\tabref}[1]{Table~\ref{#1}\xspace}

\newcommand{\validComments}{447\xspace}
\newcommand{\inspectedComments}{739\xspace}
\newcommand{\prs}{240\xspace}
\newcommand{\validPrs}{179\xspace}
\newcommand{\successRate}{10\%\xspace}    

\begin{document}

\title{Studying Quality Improvements Recommended via Manual\\and Automated Code Review}


\author{Giuseppe Crupi}
\affiliation{%
  \institution{Universit\`a della Svizzera italiana}
 \country{Switzerland}}
 \email{giuseppe.crupi@usi.ch}

\author{Rosalia Tufano}
\affiliation{%
  \institution{Universit\`a della Svizzera italiana}
 \country{Switzerland}}
\email{rosalia.tufano@usi.ch}

\author{Gabriele Bavota}
\affiliation{%
  \institution{Universit\`a della Svizzera italiana}
 \country{Switzerland}}
\email{gabriele.bavota@usi.ch}

\copyrightyear{2026}
\acmYear{2026}
\setcopyright{cc}
\setcctype{by}
\acmConference[ICPC '26]{34th IEEE/ACM International Conference on Program Comprehension}{April 12--13, 2026}{Rio de Janeiro, Brazil}
\acmBooktitle{34th IEEE/ACM International Conference on Program Comprehension (ICPC '26), April 12--13, 2026, Rio de Janeiro, Brazil}
\acmPrice{}
\acmDOI{10.1145/3794763.3794809}
\acmISBN{979-8-4007-2482-4/2026/04}

\begin{abstract}
Several Deep Learning (DL)-based techniques have been proposed to automate code review. Still, it is unclear the extent to which these approaches can recommend quality improvements as a human reviewer. We study the similarities and differences between code reviews performed by humans and those automatically generated by DL models, using ChatGPT-4 as representative of the latter. In particular, we run a mining-based study in which we collect and manually inspect \inspectedComments comments posted by human reviewers to suggest code changes in \prs PRs. The manual inspection aims at classifying the type of quality improvement recommended by human reviewers (\eg \emph{rename variable/constant}). Then, we ask ChatGPT to perform a code review on the same PRs and we compare the quality improvements it recommends against those suggested by the human reviewers. We show that while, on average, ChatGPT tends to recommend a higher number of code changes as compared to human reviewers ($\sim$2.4$\times$ more), it can only spot \successRate of the quality issues reported by humans. However, $\sim$40\% of the additional comments generated by the LLM point to meaningful quality issues. In short, our findings show the complementarity of manual and AI-based code review. This finding suggests that, in its current state, DL-based code review can be used as a further quality check \emph{on top} of the one performed by humans, but should not be considered as a valid alternative to them nor as a mean to save code review time, since human reviewers would still need to perform their manual inspection while also validating the quality issues reported by the DL-based technique. 
\end{abstract}

%
%


\maketitle

\section{Introduction} \label{sec:intro}

Code review is widely spread in both open source and industrial projects. Its benefits are well-known and include higher code quality, less likelihood of bugs \cite{mcintosh:msr2014,morales2015saner,bavota:icsme2015,caitlin:icse2018}, and knowledge transfer \cite{bacchelli2013expectations}. 
Those benefits, however, do not come from free: Developers can spend substantial amount of time reviewing code \cite{rigby:fse2013,Rigby:tosem2014,Bosu:2013}, thus increasing the costs associated to its development. 

Researchers recently proposed Deep Learning (DL)-based techniques aimed at partially automate code review, with the  motivation of saving developers' time \cite{tufano:icse2022}. The evaluations reported in the literature for these techniques are mostly based on quantitative metrics looking, for example, at the textual similarity between comments posted by human reviewers and those generated by the DL model for the same code under review. These metrics can objectively compare different approaches, but say little about their  capabilities in identifying code quality issues as done by humans. 

In this work, we aim at partially filling this gap by presenting empirical evidence about the ability of DL-based techniques to recommend the same quality improvements spotted by human reviewers. First, we mine PRs from Java and Python open source repositories hosted on GitHub. We only focus on PRs which have been subject to a code review process. We extract from the PRs a statistically significant sample of \inspectedComments reviewers' comments, manually classifying the type of quality improvement recommended by the human reviewers (if any). This phase excluded 292 reviewers' comments as false positives, namely comments not recommending any code quality improvement (\eg \emph{thanks!}). Through this process we obtained  \validComments valid comments from \validPrs PRs. We then exploit ChatGPT-4 Turbo \cite{chatgpt} as representative of DL-based code review automation techniques. Indeed, recent studies \cite{tufano:msr2024} showed that open source projects are starting adopting ChatGPT as a co-reviewer commenting on submitted PRs. ChatGPT-4 Turbo supports up to 128k tokens as input, thus allowing us to input the Large Language Model (LLM) with all changes implemented  in a PR, and ask for a code review as output. We ask ChatGPT to review the same \validPrs PRs for which we labeled at least one valid reviewer comment. Then, we manually check whether ChatGPT was able to recommend the same code quality improvements suggested by the human reviewer(s).

We found that ChatGPT tends to report a higher number of potential quality issues as compared to humans ($\sim$2.4$\times$ more). However, only \successRate of the quality improvements suggested by humans are also recommended by ChatGPT. We also show that the type of quality improvement posted by the human reviewer (\eg a refactoring recommendation) does not impact the likelihood that ChatGPT will recommend the same code change. This leads to the question of whether the additional comments generated by ChatGPT but not reflecting quality issues identified by humans are actually relevant or just noise. To answer this question, we manually inspected 292 of these comments, finding 40\% of them to actually recommend meaningful code changes, which at least should warrant a consideration by the developers. Our findings show the complementarity between manually written and AI-generated code reviews, suggesting the integration of AI-based code review as a further check after the manual inspection rather than as a replacement for the latter.
\section{Related Work} \label{sec:related}

\subsubsection*{Code Review Automation}

First attempts in the automation of code review-related tasks focused on classification problems, such as recommending the most appropriate developer(s) to review a code change \cite{thongtanunam:saner2015, xia:icsme2015, ouni:icsme2016, rahman:icse2016, zanjan:tse2016, hannebauer:ase2016, asthana:esecfse2019, jiang:jss2019, strand:icse-sep2020, pandya:esecfse2022}, or  decomposing a code change submitted for review \cite{barnett:icse2015}. 

The advent of DL in software engineering made it possible to tackle generative tasks, such as \emph{code refinement} \cite{li:fse2022, tufano:icse2021, tufano:icse2022, bensghaier:fse2024} and \emph{review comment generation} \cite{li.l:esecfse2022, li:fse2022, hong:esecfse2022, tufano:icse2021,  tufano:icse2022, lu:issre2023, bensghaier:fse2024}. 
The former consists in automatically implementing code changes recommended by the reviewers on a given piece of code. This means that the model is trained to take as input the code submitted for review ($C_s$) and a reviewer's comment in natural language ($R_{nl}$) and produce as output the revised code addressing $R_{nl}$ (\ie it must implement in $C_s$ code changes aimed at addressing $R_{nl}$). 

Concerning \emph{review comment generation}, the idea is to teach the DL model how to comment on a code change by recommending improvements in natural language, as a human reviewer would do. In this case, the model is trained to take as input $C_s$ and produce as output $R_{nl}$. This task is the focus of our paper, in which we use ChatGPT-4 Turbo \cite{chatgpt} as representative of a DL-based technique which can support \emph{review comment generation} \cite{tufano:msr2024}.

DL-based solutions have also been proposed for predicting whether a code change should be accepted or not \cite{li:fse2022,islam:ist2022}.

\subsubsection*{Empirical Studies on Code Review}

Empirical studies have been conducted to unveil the most common practices adopted in code review \cite{rigby:icse2008, rigby:fse2013, bosu:tse2017}, the kind of quality issues code review is able to catch \cite{bacchelli:icse2013,fregnan:emse2022}, and its impact on code quality \cite{mcintosh:msr2014, belleretal:msr2014, bosucarver:icse2014,bavota:icsme2015}. Human biases while reviewing peers code have also been studied \cite{Spadini:icse2020, Fregnan:fse2022}. For example, Fregnan \etal \cite{Fregnan:fse2022} showed that the order in which files are inspected during review impacts the likelihood of identifying bugs in them. Other works focused on possible strategies to improve the effectiveness of the code review process \cite{Goncalves:msr2020}, such as the integration of static analysis tools as preliminary automated code quality checks \cite{balachandran:icse2013, panichella:saner2015, beller:saner2016}.

Tufano \etal \cite{tufano:icse25} conducted a controlled experiment with 29 professional developers to study the impact of including AI-generated code reviews in the code review process, focusing in particular on the reviewer's ability to identify quality issues and the time spent to complete the review. The authors reveal that providing an AI-generated code review as a starting point to the reviewers is counterproductive, since it pushes the reviewers to only focus on the code locations pointed in the AI-generated review, overlooking quality issues in other parts of the code.

Guo \etal \cite{guo2024exploring} tested the ability of ChatGPT in supporting the previously mentioned \emph{code refinement task} (\ie automatically addressing a code review comment by implementing code changes in the submitted code). 

The authors exploit automated metrics to assess ChatGPT's performance, such as BLEU \cite{papineni:acl2002} and exact match (EM) predictions (\ie the revised code is identical to the one manually revised by the developers during the code review process). Our study focuses on the \emph{review comment generation} task,  for which automated metrics say very little about the performance of an approach. For example, EM predictions (\ie the comment generated by ChatGPT is exactly the same as the one posted by a human reviewer) might not be the best option for this task, since the same concept can be expressed using completely different wordings, thus also invalidating metrics such as BLEU. This is the reason why our study has a strong qualitative focus and completely relies on manual analysis of the review comments written by humans and generated by ChatGPT.  

Tufano \etal \cite{tufano:tse2024} performed a qualitative evaluation of three code review automation techniques proposed in the literature \cite{tufano:icse2022,hong:fse2022,li:esecfse2022}. In particular, the authors manually inspected the predictions  generated by these approaches on the test sets used in the original papers presenting them to understand the strengths and weaknesses of the three approaches. 
The work by Tufano \etal \cite{tufano:tse2024} does not focus on assessing whether the experimented approaches are actually able to identify code quality issues as human reviewers, but rather on providing a more qualitative evaluation of their predictions. Indeed, the pairs $\langle C_s$, $R_{nl} \rangle$ on which the techniques have been evaluated (\ie those present in the test sets of the papers presenting the three approaches) have been collected from open source projects by applying strong automated filters needed to reduce the cost of the training procedure. This resulted, for example, in the exclusion of $\langle C_s$, $R_{nl} \rangle$ pairs in which the concatenation of $C_s$ and  $R_{nl}$ was longer than 512 tokens \cite{tufano:icse2022}. This means, for example, that long review comments written by humans are not part of the evaluation. Similarly, the view of the experimented techniques on the code changes implemented in a PR is very limited (\ie a single impacted function or a single diff hunk), thus making unfair any comparison with human code inspections. In our work, we exploit ChatGPT-4 Turbo \cite{chatgpt} as a DL-based code review technique to compare with human reviewers, taking advantage of its ability to take as input up to 128k tokens, thus being able to inspect entire files submitted for review together with their diff. Also, we compare its predictions with all (relevant) comments written by human reviewers for the same PR, without applying any filter which can make any comparison between the ``human and the machine'' unfair.
\section{Study Design} \label{sec:design}

We aim at answering the following research question (RQ):

\emph{Which code quality improvements are recommended during manual and automated code review?}

Given a set of PRs, we characterize the code quality improvements recommended during manual and automated code inspection. In particular, for each PR, we manually inspect and classify the comments posted by human reviewers. Then, we run on the same PR ChatGPT-4 Turbo \cite{chatgpt} asking it to perform a code review. All code and data used in our study is publicly available \cite{replication}.

\subsection{Study Context}
We use GHS \cite{dabic:msr2021} to select projects for our study. GHS is a platform to sample GitHub repositories based on specific selection criteria. 

We query it to collect all Python and Java public GitHub projects having at least 10 contributors (to exclude personal projects), 500 commits (to exclude toy projects), and 100 PRs (as they are the main data used in our study). We sort the returned list of projects in descending order by number of stars and select the first 100 Python and first 100 Java projects. The first author manually inspected the 200 projects to identify and remove repositories not representing actual software systems (\eg Java tutorials showcasing design patterns implementations). This resulted in the removal of 22 projects (15 Python and 7 Java), leading to 85 Python and 93 Java repositories. Finally, we excluded all projects not having at least one commit in the past 30 days, to focus our analysis on active projects, obtaining the 46 repositories (24 Python and 22 Java) subject of our study.

Next, using the GitHub APIs, we collected for each subject project its 100 most recent PRs (based on the creation date) with their metadata. We then removed all PRs: (i) not having comments, since they have not been subject of discussion (at least, not on GitHub); (ii) opened by bots, identified as users explicitly labeled in the GitHub APIs as bots or having in their username the words ``bot'' or ``github'', since those are mostly related to small automated changes such as library updates; and (iii) not merged, since we want to focus on code changes which have been considered relevant for the project and thus, more likely to be thoroughly reviewed.

We collected all comments from the selected PRs, excluding the ones posted by the author of the PR or by bots, using the same heuristic previously described. Indeed, we are only interested in collecting reviewers' comments for our manual analysis aimed at categorizing the type of code quality improvements recommended in the review process (details in \secref{human_data_collection}). We also excluded comments containing non-ASCII characters, as an attempt to exclude non-English comments, as well as the ones linked to files which are not Java or Python (\eg configuration files such as \texttt{pom.xml}).

\begin{table}[t]
\centering
    \caption{Manually Inspected Review Comments}
    \label{tab:datasets}
    {
    \footnotesize
    \begin{tabular}{lrrr}
    \toprule
    \textbf{Language}                       & \textbf{\#Projects}  & \textbf{\# PRs} & \textbf{\# Comments}\\ \midrule
    Python & 19 & 116 &  382\\
    Java & 21 & 124 & 357\\\midrule
    Total & 40 & 240 & \inspectedComments\\
    \bottomrule
    \end{tabular}
    }
\end{table}

At the end of this cleaning process, we ended up with 1,016 comments from 279 Python PRs, and 620 comments from 168 Java PRs. We targeted the manual inspection of at least 700 review comments across the two languages, representing a statistically significant sample with a confidence level of 99\%  and confidence interval of $\pm5\%$. To select these comments, we split the PRs of each language into four quartiles based on the number of review comments they received. For example, for Java, Q$_1$= 1, Q$_2$= 2, and Q$_3$= 4. Then, for each language, we randomly selected an equal number of PRs from each quartile until we reached our target. The final datasets are summarized in \tabref{tab:datasets}. 

\subsection{Data Collection}
\label{sub:collection}
The review comments we collected have been manually inspected through the process described in \secref{human_data_collection} to identify the type of code quality improvement (if any) recommended by the reviewer. Then, the same code files subject of the human review have been provided to ChatGPT for a DL-based code review (\secref{chatgpt_data_collection}).

\subsubsection{Issues Identified by Human Reviewers}  \label{human_data_collection}

The goal of the labeling process was to assign each review comment one or more labels describing the quality improvements recommended by the reviewer. Each comment has been labeled independently by two authors (evaluators). As a starting set of labels, we used the ones defined by Tufano \etal \cite{tufano:tse2024}, which categorized the types of code changes recommended in the review comments present in the test sets of DL-based code review techniques (see \secref{sec:related} for additional details). Since these test sets have been built by filtering out instances being problematic for the DL model training/testing (\eg review comments being too long), the labels from Tufano \etal should represent a subset of the ones we will find. Thus, the evaluators defined new labels when needed. The evaluators were also in charge of discarding instances being unclear (\ie it was not possible to infer the quality improvement suggested by the reviewer) or just irrelevant for the study (\eg a ``thank you'' comment). 

The labeling process has been supported by a web application we built. The application randomly assigns the instances to label to the evaluators, ensuring that each instance is assigned to two evaluators. For each instance, the evaluator had on screen the review comment to label and the link to the commented PR (to inspect the code change when needed). The web application also managed the selection of existing labels or the definition of new ones. Newly added labels defined by an evaluator were immediately visible to the other evaluators, to promote a more consistent usage of labels and avoiding the definition of different labels being semantically equivalent (\eg \emph{extract method} and \emph{extract method refactoring}). 

Out of the \inspectedComments inspected comments, 290 (39\%) resulted in some form of conflict. While the percentage might seem high, it is worth remembering that the set of labels to use was not fixed (\ie evaluators had the chance to define their own labels), thus increasing the chances of conflicts. Also, we considered as conflicts cases in which the two evaluators assigned a different but overlapping set of labels (\eg the first evaluator assigned labels $L_1$, $L_2$, while the second only $L_1$). After solving conflicts, we labeled \validComments relevant code review comments, while 292 have been discarded. 

\subsubsection{Issues Identified by DL-based Code Review}  \label{chatgpt_data_collection}

We asked ChatGPT to review each of the code files linked to at least one of the \validComments comments we manually labeled with the type of quality improvement recommended by the reviewer, so that we can then check if the LLM was able to identify the same type of quality issue spotted by humans. We use the following prompt:\smallskip

\emph{Given the following code file ``\{codeFile\}'' which has been subject to the following code change ``\{diff\}''. Assume that you are an expert code reviewer, list any issue you see in the implemented code change, or just answer ``none'' if you feel the code change is fine. Be brief and avoid describing the implemented change.}
\smallskip

The prompt provides ChatGPT with the source code file before the changes implemented in the PR and the \emph{diff} of the implemented changes. It is important to note that a PR might involve changes to $n$ different files, and in our prompt we are only providing ChatGPT with one of the impacted files (\ie the one which was subject of a human reviewer comment we classified). 

This limits the contextual information available to the LLM, and may lead to lower chances that it identifies the same code quality issues spotted by humans. Indeed, some quality issues may be spottable only when considering a broader context including all files impacted in the PR or even the entire code base. We discuss this aspect further in \secref{sub:implications}. For now, we just anticipate that providing a broader context to ChatGPT (\eg all files impacted in the PR) actually lowers the percentage of cases in which it is able to spot the same quality issues identified by humans. This is why our main results discussion is centred around the code reviews obtained with the above-defined prompt. Such a prompt is the result of trials we made aimed at ensuring ChatGPT was understanding the required task. In \secref{sec:threats} we show that using other prompts is unlikely to change our main findings.

We stored the reviews generated by ChatGPT and manually inspected them to check whether they recommended the same code quality improvements suggested by the human reviewers. The first author inspected each of the \validComments pairs of $<$human\_comment, ChatGPT\_review$>$, assigning one of three possible tags: \emph{matched}, indicating that ChatGPT was able to recommend the same quality improvement suggested in the human comment; \emph{partially matched}, indicating that ChatGPT identified the same type of issue, in the same code location, but recommended a different solution or was quite vague about the presence of the issue (\eg the human reviewer recommended to throw an exception, while ChatGPT suggested to improve the error handling); \emph{missed}, indicating that ChatGPT did not recommend the same quality improvement of the human reviewer. Then, the second author double checked the classification for all Python instances, while the third author did the same for the Java instances. Out of the \validComments inspected pairs, for 99 (22\%) of them there was a disagreement, solved via open discussion. 

The data extracted up to this moment allows us to check the extent to which ChatGPT is able to recommend the same code quality improvements suggested by human reviewers, which is the main goal of our study. There are, however, all the unmatched ChatGPT comments (\ie those that identify quality issue not reported by the human reviewers) which are worth being investigated to understand which quality issues the LLM tend to report despite human reviewers ignoring them. Each ChatGPT review can feature several comments (on average, 7.21 comments per review in the reviews we obtained). The comments are usually organized as a bullet list, with each of them having a ``title'' (\eg \emph{Potential for NullPointerException}) and a description (\eg \emph{The \texttt{get\-Main\-Activity\-Name()} method processes nodes irrespective of whether they exist or not, particularly when accessing [...]}). In total, 1,290 comments were part of the generated ChatGPT reviews, 1,189 of which did not match a human reviewer comment. Since the manual effort required to analyze and classify all of them to unveil the reported type of issue would be excessive, we applied a semi-automated process. First, we parse the code reviews to extract the ``title'' of each reported issue. Then, we manually inspected the list of unique titles clustering them into ``issue type'' categories. A subset of the code review generated by ChatGPT (74) did not follow the above-described ``bullet list template''. For those, we manually read the review and split it into comments, which have then been classified as the ones part of the bullet lists.  These manual analyses have been performed by one of the authors and double checked by a second author. 

The above-described analysis of unmatched ChatGPT's comments only identifies code quality improvements often reported by ChatGPT and ``ignored'' by humans. However, it does not assess whether the comments made by ChatGPT and not matched to any human-written comment are actually meaningful. To this aim, the first two authors independently inspected 292 of the 1,189 unmatched comments, representing a statistically significant sample (95\%$\pm$5\%). The initial idea was to classify each unmatched comment as \emph{meaningful} (\ie the quality issue spotted by ChatGPT is actually in the code and has been missed in the manual review process) or \emph{not meaningful} (\ie given the implemented change, the ChatGPT's comment is not relevant). The two authors started by classifying the first 30 comments, and then had a meeting with the third author to discuss about the assigned labels and the (9) conflicts they had. After the meeting, it was clear that (i) a boolean classification (\ie \emph{meaningful} \emph{vs} \emph{not meaningful}) was inappropriate, since there were some generic comments posted by ChatGPT which, per se, were not wrong (\emph{not meaningful}) but just quite generic and not actionable; and (ii) it was difficult for us to assess the relevance of some comments (as non-developers of the projects on which the PR were opened). For this reason, we revised the possible categories to use, adding on top of 
\emph{meaningful} and \emph{not meaningful} also: (i) \emph{generic} $\rightarrow$ \emph{asks to double-check the change}, indicating comments in which ChatGPT asks the developers questions which can be abstracted as ``are you sure this is really what you want to do?''; (ii) \emph{generic} $\rightarrow$ \emph{asks if additional changes are needed}, related to comments in which ChatGPT does not spot any quality issue, but suggests potential additional changes that the developer may want to implement; (iii) \emph{generic} $\rightarrow$ \emph{summarizes the change}, being comments in which ChatGPT just summarizes the implemented change; and (iv) \emph{not evaluated}, representing cases in which we felt our programming expertise was not sufficient to judge the relevance of the ChatGPT's comment, since they would require deep knowledge of the subject project and of the intentions behind the implemented change. As a guideline, we decided to be conservative, and to assign \emph{not evaluated} in all cases in which we were doubtful.

Once completed the manual inspection, the first two authors had a meeting to solve conflicts, present for $\sim$20\% of the inspected instances. On 16 of them (5\%), they did not manage to find an agreement, with the conflict which was solved by also involving the third author.

\subsection{Data Analysis}

We present a taxonomy (\figref{fig:taxonomy}) showing the quality improvements recommended by humans and by ChatGPT, as representative of DL-based code review automation techniques. The taxonomy shows, for each type of quality improvement (\eg \emph{improve assert statement}) recommended at least once by a human across the \validComments comments we labeled, the percentage of times that ChatGPT reported it as well. We show both the cases that we tagged as \emph{matched} and as \emph{partially matched}. 

Concerning the ChatGPT comments that were unmatched, we show the distribution of types of code quality improvements recommended by ChatGPT and not by humans. Also, we report statistics about the meaningfulness of these unmatched comments accompanied by concrete qualitative examples.

%


\section{Results Discussion} \label{sec:results}

\begin{figure*}
	\centering
	\includegraphics[width=0.98\linewidth]{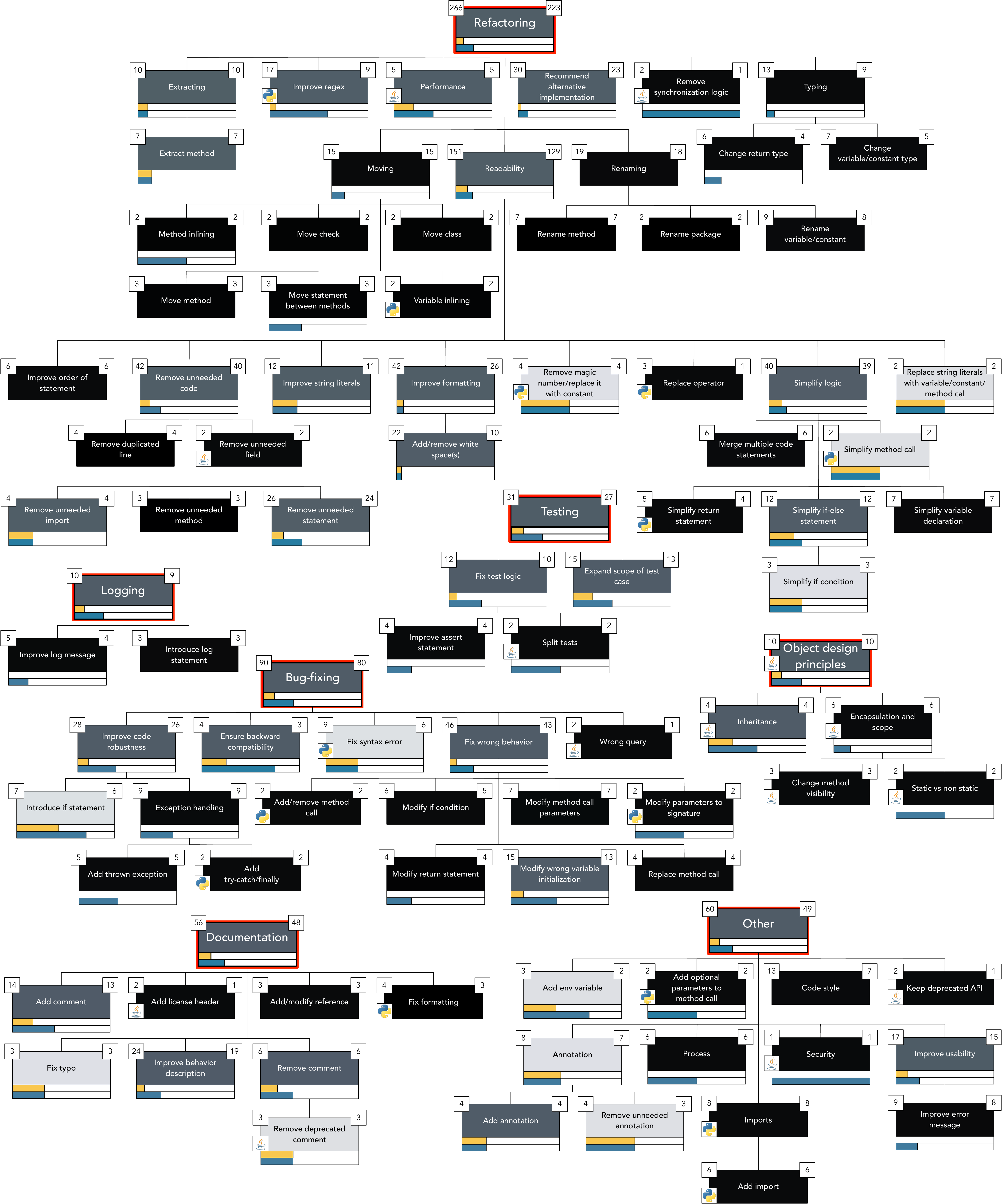}
	\caption{Taxonomy of quality improvements recommended by humans. Bars below each type of quality improvement indicates the percentage of cases in which ChatGPT recommends exactly the same code change (yellow) or at least a related one (blue).}
	\label{fig:taxonomy}
\end{figure*}

\figref{fig:taxonomy} depicts the taxonomy of code quality improvements recommended by human reviewers in the set of \validComments reviewers' comments we labelled. The categories in the taxonomy are hierarchically organized, with the child categories being specializations of their parent (\eg \emph{Refactoring} $\rightarrow$ \emph{Readability} $\rightarrow$ \emph{Improve Formatting} $\rightarrow$ \emph{Add/remove white spaces}). Root categories are characterized by a red border. Each category is labelled with two numbers: The one on the top-right corner represents the total number of PRs in which we found at least one reviewer's comment recommending the related quality improvement. The one on the top-left corner, instead, reports the number of comments labelled with that category. By construction, the number of comments is always higher or equal than the number of PRs (\ie a PR may feature more than one comment recommending the same type of code quality improvement). For example, 223 PRs featured at least one of the 266 comments recommending a refactoring operation. Worth noticing is that the sum of the instances in a parent category does not always match the sum of the instances in its child categories. This is due to the fact that during our manual inspection we assigned the most specific label we could derive for a given review comment: In some cases, the comment was so detailed that we could infer a quite specific label (\eg ``\emph{Please remove the extra spaces before statement 3}'' $\Longrightarrow$ \emph{Refactoring} $\rightarrow$ \emph{Readability} $\rightarrow$ \emph{Improve Formatting} $\rightarrow$ \emph{Add/remove white spaces}) while for other comments this was not the case (\eg ``\emph{Take care of better formatting this method}'' $\Longrightarrow$ \emph{Refactoring} $\rightarrow$ \emph{Readability} $\rightarrow$ \emph{Improve Formatting}). 
\figref{fig:taxonomy} also shows the extent to which ChatGPT was able to recommend the same code quality improvements in the same reviewed PR. In particular, \figref{fig:taxonomy} adopts the following color schema for the background of each category:

\underline{Black}: ChatGPT never managed to recommend the same code quality improvement suggested by the human reviewers for exactly the same code location. For example, none of the 7 rename method refactorings recommended by human reviewers in our dataset was also suggested by ChatGPT.

\underline{Dark grey}: ChatGPT was able to recommend the same code quality improvement suggested by the human reviewers for exactly the same code location in less than 33\% of cases (but at least once). See \eg the \emph{Testing} $\rightarrow$ \emph{Expand scope of test case} category.

\underline{Light grey}: ChatGPT was able to recommend the same code quality improvement suggested by the human reviewers for exactly the same code location in more than 33\% but less than 66\% of cases. See \eg the \emph{Documentation} $\rightarrow$ \emph{Fix typo} category.

The exact percentage of cases in which ChatGPT managed to recommend the same quality improvement suggested in the human comment is represented by the yellow bar below the grey-colored categories. This bar is not present for the black-colored categories since it would be trivially 0\%. As per the blue bars, they show instead the percentage of cases in which ChatGPT managed to recommend the same quality improvement suggested in the human comment or, at least, the same type of issue (while recommending a different solutions). In other words, referring to the terminology used in \secref{sub:collection}, the yellow bar is the percentage of $<$human\_comment, ChatGPT\_review$>$ pairs to which we assigned a \emph{matched}, while the blue bar is the percentage of pairs to which we assigned a \emph{matched} or a \emph{partially matched}.

Finally, categories which are language-specific (\ie reviewers' comments were all related to one of the two languages we considered) feature the icon of the language in their bottom-left corner. If no icon is present, the category is relevant for both languages.

Before commenting on the different root categories, there is one clear message that stands out from our taxonomy: \emph{Very rarely ChatGPT is able to recommend the same code quality improvements suggested by human reviewers given the same code to review}. Indeed, out of the \validComments valid reviewers' comments that contribute to our taxonomy, only for 45 of them (\successRate) ChatGPT managed to match the human reviewer recommendation. If we consider both the \emph{matched} and \emph{partially matched} comments, this number goes up to 103 (23\%). 

The second take-away message of our taxonomy concerns the homogeneous results observed across all sub-trees: We did not observe any relevant difference in the success rate of ChatGPT across the different sub-trees representing ``families'' of code change types (\eg \emph{Refactoring}, \emph{Bug-fixing}). This is shown by the dark grey color assigned to all root categories (\ie percentage of matched comments lower than 33\%). In the following we discuss some interesting examples, focusing on the most popular categories in our taxonomy.

Not surprisingly, \emph{Refactoring} is the main sub-tree in our taxonomy and groups together all code quality improvements not aimed at changing the behavior of the code but rather at boosting some non-functional requirements such as \emph{performance}, \emph{readability}, and \emph{maintainability}. It features popular refactoring operations including, for example, \emph{extract method}, \emph{move method}, and \emph{rename variable/constant}. 
In line with the overall trend, a very small percentage of the 266 refactoring-related comments in our dataset are matched by ChatGPT (8\%). Interestingly, this trend is also valid for refactoring operations not really requiring a strong knowledge of the code base or a large amount of contextual information (see \eg \emph{Rename method}, \emph{Rename variable/constant}). 

A concrete example is the PR 24387 from Apache Flink \cite{24387}. The code diff provided to ChatGPT for review features, among other modifications, the following change:

\begin{lstlisting}[language=diff]
@@ -497,7 +497,7 @@ 
-  Time DEFAULT_PERIODIC_COMPACTION_TIME = Time.days(30);
+  Duration DEFAULT_PERIODIC_COMPACTION_TIME = 
+  Duration.ofDays(30);
\end{lstlisting}

Looking at this change, the reviewer recommended to rename \ttfamily DEFAULT\_\-PERIODIC\_\-COMPACTION\_\-TIME\normalfont~to \ttfamily DEFAULT\_\-PERIODIC\_\-COM\-PACTION\_\-DURATION \normalfont so that the identifier can better match the semantic of the implemented code. ChatGPT did not recommend any change when reviewing this code. 

Other cases from the refactoring taxonomy in which we expected a better ``success rate'' from ChatGPT were those related to the 42 comments by the human reviewers suggesting the \emph{Removal of unneeded code}. For example, the PR 8999 of Jenkins \cite{8999} features a diff in which the following \ttfamily null\normalfont~check has been added:

\begin{lstlisting}[language=diff]
@@ -245,9 +245,9 @@ 
+  if ((userIdEnv != null && !userIdEnv.isBlank()) && [...]
\end{lstlisting}

With the (already existing) line following the \ttfamily if\normalfont~statement left unchanged:

\begin{lstlisting}[language=diff]
auth = StringUtils.defaultString(userIdEnv).[...]
\end{lstlisting}

The reviewer suggested to ``\emph{remove [...] StringUtils.defaultString}'' since they ``\emph{already know that the string was not null and not empty}'' thanks to the newly added \ttfamily if\normalfont~statement. ChatGPT missed the recommendation of this refactoring, despite having both the original file as input (thus, the unchanged statement) as well as the implemented diff. In other cases, ChatGPT could not instead recommend the removal of an unneeded statement since this would have required knowledge of the whole code base, \eg a \ttfamily null\normalfont~check was not needed for a parameter in function \ttfamily A\normalfont~ since the callee of \ttfamily A\normalfont~ already implemented this check (but the callee code was not part of ChatGPT's input, since not impacted by the PR). We further investigate the lack of additional coding context as a possible ``failure cause'' for ChatGPT in \secref{sub:implications}.

Despite the overall ``negative'' trend, there are 47 quality improvements in the refactoring tree which are recommended both by ChatGPT and by the human reviewers. For example, in PR 5072 of Apache Apollo \cite{5072}, the diff featured the following statement:

\begin{lstlisting}[language=diff]
-  .add("parentClusterId", parentClusterId)
+  .add("parentClusterId", parentClusterId)
+  .add("comments", comment)
\end{lstlisting}

\noindent with the reviewer recommending to rename the \ttfamily comments\normalfont~literal to \ttfamily comment\normalfont. Similarly, ChatGPT commented ``\emph{When using the plural form `comments', it may suggest multiple comments which do not align with the actual single `comment' attribute}''. 

Moving to the second-largest sub-tree (\ie \emph{Bug fixing}), we found ChatGPT to successfully predict a few needed checks for  \ttfamily null\normalfont~values (\emph{Bug fixing} $\rightarrow$ \emph{Improve code rebustness} $\rightarrow$ \emph{Introduce if statement}). A concrete example is PR 5030 from Apache Apollo \cite{5030}, in which the human reviewer commented the lack of ``\emph{handling mechanisms for null value scenarios}'' and, similarly, ChatGPT commented: ``\emph{In the `findMethod' method, when a `NoSuchMethodException' is caught, `null' is returned. This might lead to NullPointerException later if the result of `findMethod' is not properly checked before being used}''. 

Also the \emph{Bug fixing} sub-tree, as already observed for the \emph{Refactoring} one, features ChatGPT's failure cases attributable to lack of context, \eg category \emph{Bug fixing} $\rightarrow$ \emph{Fix wrong behavior} $\rightarrow$ \emph{Replace method call}. The latter usually requires a complete view of the system and/or of the libraries it depends upon. Others, instead, highlight again the limits of the LLM in identifying quality issues, such as the cases in which the human reviewers spot bugs in the \ttfamily if\normalfont~conditions of the implemented code (\emph{Bug fixing} $\rightarrow$ \emph{Fix wrong behavior} $\rightarrow$ \emph{Modify if condition}), while ChatGPT misses all of them. 

The \emph{Documentation} sub-tree provides the chance to discuss an interesting pattern we observed in ChatGPT's reviews, namely very general comments which are provided in almost any review it generates. A concrete example relates to PR 44045 from the Ray project \cite{44045a}, in which the human reviewer recommends to rephrase a comment documenting newly added code, by providing an explicit suggestion of how the new comment should be written (\emph{Documentation} $\rightarrow$ \emph{Improve behavior description}). ChatGPT commented the change by writing: ``\emph{The added blocks of code lack detailed comments explaining the rationale, especially in complex conditional logic areas. Commenting intent and logic is crucial for long-term maintainability}''. ChatGPT was able to identify the quality issue but, differently from the human reviewer, did not recommend how to fix it. This is one of ChatGPT's comments labeled as \emph{partially matched}.

For the \emph{Logging}, \emph{Object design principles}, and \emph{Other} subtrees, we reiterate the inability of ChatGPT to recommend the same quality improvements suggested by humans (10\% of human comments matched for all three categories). 

We conclude by discussing one example from the \emph{Other} $\rightarrow$ \emph{Add annotation} category which shows how ChatGPT can be subject to artificial hallucination. While reviewing the code changes in PR 10146 of the Thingsboard project \cite{10146}, ChatGPT commented: ``\emph{The `save\-Device\-With\-AccessToken' method internally calls `doSaveDevice', which is already annotated with `@Transactional'. Adding `@Transactional' to `save\-Device\-With\-AccessToken' may be redundant if `doSaveDevice' already appropriately handles transaction boundaries}. However, the \ttfamily do\-Save\-Device\normalfont~was not annotated with \ttfamily @Transactional\normalfont. This final example stresses the importance of double-checking ChatGPT's review comments, highlighting its potential only as co-reviewer of a human developer.

\begin{table*}
\centering
\footnotesize

    \caption{Classification of ChatGPT's Unmatched Comments}
    \label{tab:unmatched}
    {
    \begin{tabular}{m{3.4cm}rrrm{10cm}}
    \toprule
    \textbf{Issue}                       & \textbf{\#Total}  & \textbf{\#Java} & \textbf{\#Python} & \textbf{Example} \\
    \textbf{Type}                       & \textbf{PRs}  & \textbf{PRs} & \textbf{PRs} & \textbf{} \\
    \midrule
    Bugs, Robustness, and Error Handling & 126 (70\%) & 60 & 66 &  \emph{Potential Issue with Database Setup and Cleanup: The SQL scripts are run before and after the test which is good for isolating tests, but if any test fails, ensuring that the cleanup SQL is always executed regardless of the test outcome should be verified to avoid leaving the database in an altered state.}\\\midrule
    
	Documentation & 68 (38\%) & 32 & 36 &   \emph{Potentially Misleading Comments: The comment ``ms'' is left somewhat ambiguously above the `writerRetryInterval` field. It would be clearer if it is directly included in the comment block for the field itself, e.g., ``Retry interval in milliseconds.''} \\\midrule
	
	Performance, Memory, and Multithreading & 65 (36\%) & 37 & 28 &   \emph{Thread Safety and Concurrency: The handling of cache invalidation and updates, especially in concurrent environments (if applicable), should ensure that changes do not impact ongoing cache retrievals in a way that could cause inconsistencies or race conditions.} \\\midrule
	
	Cloned, Deprecated, Unneeded, and Commented-out Code & 61 (34\%) & 28 & 33 &   \emph{Redundant Import: The class imports both `com.alibaba.csp.sentinel.util.StringUtil' and `org.apache.dubbo.common.utils.StringUtils', which both provide similar functionality like checking for empty strings. Using one consistently would reduce redundancy.}  \\\midrule
	
	Software-specific & 47 (26\%) & 19 & 28 &   \emph{GL API Version Handling:
  The code change introduces calls to `GL31', but there's not enough guarding against potential use on systems where OpenGL ES 3.1 is not available, even though there's a conditional check on `Gdx.graphics.isGL31Available()'. Consistent and comprehensive compatibility checks are crucial.}  \\\midrule
	
	Testing & 47 (26\%) & 20 & 27 &   \emph{Inadequate Mock Testing: The use of `when(...).thenReturn(...)' is good, but it's important to ensure that these mocks are verified with `verify(...)' to ensure that the configurations are called as expected.}  \\\midrule
	
	Logging and Error/Testing Messages & 41 (23\%) & 20 & 21 &   \emph{Logging: More detailed logging, especially in error cases within `handleMessage' and during cache loading (in Catch blocks or after significant operations), could be beneficial for tracing and debugging production issues.}  \\\midrule

	Code Design Principles & 38 (21\%) & 19 & 19 &   \emph{Unnecessarily Strong Coupling: By using instances of `RetryableMapWriter' and `RetryableMapWriterAsync' directly in the `writer' and `writerAsync' methods, it tightly couples `MapOptions' with these retry implementations. Consider providing a way to configure this behavior externally instead of hardcoding it.}  \\\midrule
		
	Hardcoded Data & 34 (19\%) & 14 & 20 &   \emph{Hardcoding in Test: The hardcoded values such as the IDs (e.g., `Assert.assertEquals(6, history.getId())') make the test brittle and dependent on specific dataset setups which may not be robust against data changes. It's generally better to use more dynamic assertions based on the manipulated data.}  \\\midrule

	Understandability & 33 (18\%) & 19 & 14 &   \emph{Code Complexity: The `createOrcPageSource' method is very lengthy and does a lot of different things. For better maintainability, it could be refactored into smaller, more focused methods. This would also improve testability.}  \\\midrule
	
	Formatting, Style, and Identifiers & 32 (18\%) & 18 & 14 &   \emph{Parameter Naming: The Boolean parameter `useGroupAndVersion' in methods might be more cleanly named to denote its boolean nature, like `isGroupAndVersionUsed'.}  \\\midrule	
	
	Build Process, Dependencies, and Configurations & 23 (13\%) & 10 & 13 &   \emph{Configuration Properties Dependency: The code relies on Hibernate's `SessionFactoryImplementor' to get database dialect information, which adds a dependency on Hibernate being correctly configured and available. This dependency might not be clear from the configuration files alone.}  \\\midrule
	
	Behavioral Changes & 17 (9\%) & 12 & 5 &   \emph{Potential Impact on Existing Functionality: The change includes a switch for asynchronous writers to a retry mechanism. If existing clients rely on the previous behavior where `writerAsync' directly assigns the writer with no retry logic, this might introduce unexpected behavior.}  \\\midrule
	
	Maintenance & 15 (8\%) & 6 & 9 &   \emph{Logic separation: It might be cleaner to separate handling the custom, autoscale, and range modification logic into separate methods for clarity and maintainability.}  \\\midrule
	
	Clarification Questions & 10 (6\%) & 8 & 2 &   \emph{Default Values for New Fields: Fields like `buildInfoRevision', `buildInfoBranch', [...] have been initialized with empty strings. It is unclear whether empty defaults are helpful or if they should have more meaningful defaults or validations to check if these fields are correctly populated when necessary.}  \\\midrule
	
	Security & 9 (5\%) & 3 & 6 &   \emph{Security \& Permissions: Ensure there are appropriate security checks to verify that the operator has the necessary rights to create apps. This is crucial to prevent unauthorized app modifications.}  \\
    \bottomrule
    \end{tabular}
    }
\end{table*}

\subsection{Unmatched ChatGPT's Comments} 
\label{sub:unmatched}

\tabref{tab:unmatched} reports the categorization we performed for the 1,189 ChatGPT's comments that we did not match (not even \emph{partially}) to any human comment. In this case, rather than classifying the type of quality improvement recommended by ChatGPT, it was more straightforward to classify the type of quality issue identified since, as discussed before for the \emph{Documentation} sub-tree, several of the comments generated by ChatGPT are quite general and do not recommend a specific action. \tabref{tab:unmatched} shows: The type of quality issue reported by ChatGPT (column ``Issue Type''); the total number (and percentage) of PRs in which one or more comments of this category were posted by ChatGPT but not by the human reviewer (column ``\#Total PRs''), together with a breakdown across the two languages we considered (columns ``\#Java PRs'' and ``\#Python PRs''). Finally, one concrete example of comment belonging to each issue type is reported (column ``Example'').

First, it is worth mentioning that 142 out of the \validPrs PRs subject of our study (79\%) feature at least one ChatGPT's comment which cannot be matched to any of the comments posted by the human reviewer(s). Most of these unmatched comments relate to \emph{Bugs, Robustness, and Error Handling}, followed by \emph{Documentation} issues, and suboptimal implementation choices affecting \emph{Performance, Memory, and Multithreading} or resulting in \emph{Cloned, Deprecated, Unneeded, and Commented-out Code} (see \tabref{tab:unmatched}).

\tabref{tab:relevance} provides some hints into the meaningfulness and relevance of these unmatched comments. For 23 out of the 292 we inspected, we were not able to assess the meaningfulness of the comment, assigning the ``\emph{not evaluated}'' label for the reasons explained in \secref{chatgpt_data_collection}. As for the remaining, we found 115 (43\%) to be \emph{meaningful}, thus reporting code quality issues potentially useful for developers. 

An example of ChatGPT comment labeled as meaningful relates to the PR 5262 of the Redisson project \cite{5262}. ChatGPT's comment relates to the following diff:
\begin{lstlisting}[language=diff]
+ catch (NoSuchFieldException | IllegalAccessException ignored) {
+ log.warn("ProtobufCodec failed to retrieve ..."); }
\end{lstlisting}

The LLM's suggestion aims at simplifing the debugging process: \emph{``Exceptions are caught and ignored in the static initializer block. It logs a warning but it might be more appropriate to also log the exception itself for a clearer debugging process''}. Another interesting case concerns a bug spotted by ChatGPT but missed in the human review \cite{9724}. The diff of the PR adds the following lines:

\begin{lstlisting}[language=diff]
+ if len(renewed_domains_str) > 16_000:
+  logger.warning("Limiting RENEWED_DOMAINS 
+                  environment variable to 16k characters")
+  renewed_domains_str = renewed_domains_str[:16_000]

+ if len(failed_domains_str) > 16_000:
+  logger.warning("Limiting FAILED_DOMAINS
+                  environment variable to 16k characters")
+  renewed_domains_str = failed_domains_str[:16_000]
\end{lstlisting}

ChatGPT commented:  \emph{``In run\_saved\_post\_hooks, there is a probable error where renewed\_domains\_str is assigned twice in the conditional blocks without ever using failed\_domains\_str''}. The code aims at truncating \ttfamily renewed\_domains\_str~\normalfont and \ttfamily failed\_domains\_str~\normalfont in case they exceed 16k characters. However, the second \ttfamily if~\normalfont statement (handling \ttfamily failed\_domains\_str\normalfont) mistakenly assigns the truncated version of \ttfamily failed\_domains\_str~\normalfont to \ttfamily renewed\_domains\_str\normalfont.

In contrast, ChatGPT also generated 68 \emph{not meaningful} comments among the ones we evaluated (25\%). An example of \emph{not meaningful} ChatGPT comment concerns the following diff \cite{44045b}: 
\begin{lstlisting}[language=diff]
- logger.info("Autoscaling ...")
if new_num > old_num:
+  logger.info("Upscaling ...")
elif new_num < old_num:
+  logger.info("Downscaling ...")
\end{lstlisting}

The LLM's comment states: \emph{``the new log statements added seem superfluous [...]. Having both could lead to duplicate messages, making logs more difficult to read''}. However, the two log statements are mutually exclusive since appearing in different branches of the \ttfamily if\normalfont. 

Finally, worth commenting is the substantial number of \emph{generic} comments that ChatGPT generates, which represent 32\% of the unmatched comments we evaluated. Based on our experience, these comments are likely to just represent a further burden on the developers' shoulders due to their lack of actionability (\eg ``\emph{The addition to log a warning when backoff has been repeating for a significant amount of time [...] is sensible. It offers a diagnostic insight into potential prolonged difficulties in scheduling requests [...]}'').

\begin{table}[t]
\footnotesize
\centering
		\caption{Relevance of the Unmatched ChatGPT's comments}
		\label{tab:relevance}
		{
		\begin{tabular}{lr|rr}
		\toprule
		\textbf{Category}  & \textbf{Total} & \textbf{Java}  & \textbf{Python} \\ \midrule
		\emph{Meaningful} & 115 & 51 & 64 \\
		\emph{Not meaningful} & 68 & 35 & 33 \\
		\emph{Generic} $\rightarrow$ \emph{asks to double-check the change} & 34 & 17 & 17 \\
		\emph{Generic} $\rightarrow$ \emph{asks if additional changes are needed} & 44 & 23 & 21 \\
		\emph{Generic} $\rightarrow$ \emph{summarizes the change} & 8 & 5 & 3 \\ \midrule
		\emph{Not evaluated} & 23 & 15 & 8 \\
		\bottomrule
		\end{tabular}
		}
\end{table}

\subsection{Lessons Learned}
\label{sub:implications}
\emph{There is little overlap between the code quality improvements recommended by human reviewers and by ChatGPT}. ChatGPT managed to match only \successRate of the \validComments comments posted by human reviewers in the \validPrs PRs subject of our study. Still, there is value in the ChatGPT's comments which spot quality issues not reported by humans. Indeed, we estimate $\sim$40\% of these comments to be actually valuable. Overall, our findings suggest the suitability of ChatGPT as a co-reviewer performing further checks on the code change and highlights the strong need for humans in the loop during the review process (due to quality issues missed by the LLM).

\emph{We expect ChatGPT to increase the review cost rather than reducing it}. One of the motivations for automating code review via DL models is to save developers' time \cite{tufano:icse2021}. Based on the large number of ChatGPT's comments which do not mirror any recommendation posted by human reviewers, including ChatGPT as a co-reviewer is expected to increase the review cost (time), since the human reviewers will need to validate ChatGPT's comments and, when appropriate, consider proper actions aimed at addressing them. For the reasons previously explained, by no means this is guaranteed to be a losing game: the higher cost payed at review time will likely be compensated by a higher code quality and, possibly, a lower maintenance/evolution cost in the future.

\emph{When using ChatGPT as reviewer, the contextual information provided as input may play a role in its ability to spot quality issue}. The lack of context was one of the reasons why ChatGPT could not match some of the human reviewers' comments. This may suggest that the prompt we used in our experiment likely penalized ChatGPT, making impossible for the LLM to spot the same quality issues identified by humans. To assess the extent to which this was the case, we experimented with a different prompt providing much more context to ChatGPT. In particular, the new prompt (available in our replication package \cite{replication}) includes (i) the title and description of the PR; (ii) all files impacted by the PR with their corresponding diff (\ie the implemented code changes); and (iii) all files imported by the files impacted by the PR. Before asking ChatGPT to perform the code review for a PR$_i$, we used OpenAI's Tiktoken \cite{tiktoken} to compute the number of tokens needed to represent the above-listed contextual information for PR$_i$. 

We found that for 24 out of 179 PRs (8 Java and 16 Python) it was not even possible to include all files impacted by the PR (without their imports) in the prompt. Since the goal of this analysis was to check the impact of a more comprehensive context on the code reviews generated by ChatGPT, we decided to discard these PRs. For the remaining 155 PRs, we built a prompt featuring all the files impacted in the PR and as many ``imported files'' as possible up to the maximum number of tokens accepted by the LLM. For 119 PRs we were able to fit all the imported files, while for the remaining 36 we decided to prioritize the imports from the PR-impacted files which experienced more code changes in the context of the PR (according to the sum of added and deleted lines). 

Once obtained the 155 ChatGPT reviews, we manually inspected them with the goal of matching ChatGPT's comments to the human-written comments (exactly same procedure explained in \secref{chatgpt_data_collection}). We found out that the additional context provided in the augmented prompt reduced the ability of ChatGPT to mirror the human's comments. Indeed, we went from a 10\% match achieved with our original prompt, to a 4\% with the augmented prompt (from 23\% to 6\% when also considering the \emph{partially matched} comments). Also, there was a stark increase of the PRs for which ChatGPT approved the implemented changes without pointing to any issue: this happened for 56\% of PRs with the augmented prompt, as compared to the 4\% of PRs for the original prompt. Remember that these are all PRs in which human reviewers found at least one quality issue to be fixed. 

A question may still arise about whether the ChatGPT's unmatched comments generated with the augmented prompt are of higher quality as compared to those generated with the original prompt. To answer this question, we replicated the analysis described in \secref{chatgpt_data_collection} in which we classify unmatched comments as: \emph{meaningful}, \emph{not meaningful}, \emph{generic} $\rightarrow$ \emph{asks to double-check the change}, \emph{generic} $\rightarrow$ \emph{asks if additional changes are needed}, \emph{generic} $\rightarrow$ \emph{summarizes the change}, or \emph{not evaluated}. Since for 56\% of PRs ChatGPT did not output any recommendation in its code reviews, the absolute number of unmatched comments is substantially lower in this scenario (384 in total, as compared to the 1,189 obtained with the original prompt). We manually classified 194 of these unmatched comments, randomly selected to form  a 95\%$\pm$5\% significant sample. Once excluded the 21 for which we did not manage to assess their meaningfulness (\ie labeled with \emph{not evaluated}), we observed a slight drop in the quality of the ChatGPT's comments as compared to the comments generated with the prompt featuring less contextual information. Indeed, 31\% were classified as \emph{meaningful} and 48\% as \emph{not meaningful}, (as compared to the 43\% of \emph{meaningful} and 25\% of \emph{not meaningful} of the previous prompt). The remaining ones were distributed in the \emph{generic} categories: 11\% as \emph{generic} $\rightarrow$ \emph{asks to double-check the change}, 9\% as \emph{generic} $\rightarrow$ \emph{asks if additional changes are needed},  and 1\% as \emph{generic} $\rightarrow$ \emph{summarizes the change}.

In summary, while the context provided in the prompt certainly plays a major role in the code review output obtained via ChatGPT, we cannot attribute the low percentage of matched human comments nor the overall quality of the additional unmatched ChatGPT's comments to the lack of contextual information.

\emph{ChatGPT's comments often lacks actionability and artificial hallucination in code review is possible as for most applications of DL models to generative problems}. 

We found several of the ChatGPT comments which were not matched to human comments to be quite general and pointing to ``potential'' issues. The value of these comments from the practitioners' perspective is likely to be low. 

Indeed, while some of them may lead to a further check of specific parts of the change others may just be considered a waste of time. Studies focusing on this specific aspect are needed to better investigate the phenomenon. Finally, practitioners using ChatGPT as (co-)reviewer must be aware of the possibility of artificial hallucination, with wrong/misleading comments posted in some cases.
\section{Threats to Validity} \label{sec:threats}

\textbf{Construct validity.} To collect comments from human reviewers, we mined PRs from GitHub repositories. To mitigate imprecisions in our datasets, we manually analyzed a sample of the extracted PRs and verified that the collected data was correct, especially the linking between the comment and the code line to which it referred.

\textbf{Internal validity.} The taxonomy of recommended code quality improvements we built, the mapping between humans' and ChatGPT's comments, and the assessment of the meaningfulness of the unmatched ChatGPT comments might be influenced by subjectiveness issues in our manual labeling. To at least mitigate this threat, each manually inspected artifact has been assigned to at least two researchers, with open discussions started in case of disagreement. Despite this process, we acknowledge possible imprecisions. 

Our findings are also likely affected by the specific prompt we provided to ChatGPT. Indeed, the major impact that prompting has on the output generated by LLMs has been well-documented in the literature \cite{mastropaolo:icse2023}. The prompt we defined was based on a trial and error procedure during which we mostly made sure that ChatGPT was understanding the task to be performed. We are confident this was the case. As a further check, we took three categories from our taxonomy for which ChatGPT always failed in recommending the same code quality improvement suggested by humans (\ie \emph{Refactoring} $\rightarrow$ \emph{Renaming} $\rightarrow$ \emph{Rename method}, \emph{Other} $\rightarrow$ \emph{Improve usability} $\rightarrow$ \emph{Improve error message}, and \emph{Bug-fixing} $\rightarrow$ \emph{Fix wrong behavior} $\rightarrow$ \emph{Modify if condition}) and created specialized prompts explicitly pointing ChatGPT to the code quality issue we were looking for in that specific PR (\ie the one spotted by the human reviewer). For example, in the case of the \emph{Rename method} category, we provided the following prompt as input:

\emph{Given the following code file ``\{codeFile\}'' which has been subject to the following code change ``\{diff\}''. Assume that you are an expert code reviewer, check if the names assigned to the impacted functions are meaningful and appropriate. If yes, just output ok, otherwise, suggest new function names. Be brief and avoid describing the implemented code change.}

For all three categories, ChatGPT always outputted ``ok'', again not being able to recommend the same code quality improvement suggested by human reviewers despite the further hint provided in the prompt. 

Thus, while we acknowledge the role that the used prompt may have played in our study, we are quite confident that our main findings would hold even exploiting different prompts. 

\textbf{External validity.} We only considered reviews done in the context of PRs submitted for Java and Python projects, since these are the languages for which all three authors have strong expertise and felt confident in performing the required qualitative analyses. While this limits the generalizability of our results, it is worth noticing that we did not find any substantial difference in findings across the two languages.

Similarly, we only considered ChatGPT as representative of DL-based code review techniques. Such a choice was mainly driven by recent work in the literature showing ChatGPT's adoption as ``co-reviewer'' in some open source projects \cite{tufano:msr2024}. Studies adopting other LLMs or DL models specifically trained for the code review task could help in corroborating or contradicting our findings. 

Finally, the generalizability of our findings is also  limited by the number of comments we considered (\validComments) and, as a consequence, by the number of ChatGPT's reviews we analyzed. However, the amount of required manual analyses required by our study called for capping the amount of data in some way. As for any purely qualitative study, future work can help in improving the generalizability of our findings. 
\section{Conclusion and Future Work} \label{sec:conclusion}

We investigated the extent to which a LLM, and in particular ChatGPT, is able to recommend the same code quality improvements suggested by humans when reviewing the same code change. Our study required the manual inspection of \inspectedComments reviewers' comments spread across \prs PRs opened in Java and Python projects hosted on GitHub. We managed to classify the code quality improvement suggested by the human reviewers for \validComments of these comments (belonging to \validPrs PRs). 
We then asked ChatGPT to generate a review for the same \validPrs PRs, checking whether it was able to spot the same code quality improvements suggested by the human reviewers. 

Our findings show that only \successRate of the reviewers' comments are ``mirrored'' by ChatGPT. Also, the LLM tends to spot a high number of quality issues do not reported by the human reviewers. Interestingly, $\sim$40\% of them point to meaningful code quality issues which may help in further improving the code review process after a manual inspection has been performed. 

As part of our future work  we plan to run controlled experiments aimed at investigating how human and DL-based reviewers can  cooperate to maximize the effectiveness of the review process. Also, we aim at improving the generalizability of our findings by replicating our study with DL-based techniques specifically trained for the automation of code review (see \eg \cite{li.l:esecfse2022, li:fse2022, hong:esecfse2022,   tufano:icse2022}) as well as by extending the subject projects and languages.

All the data and code used in our study is publicly available \cite{replication}.


\section*{Acknowledgment}
We acknowledge the financial support of the Swiss National
Science Foundation for the PARSED project (SNF Project No.
219294).

\bibliographystyle{ACM-Reference-Format}
\bibliography{main}

\end{document}